\documentstyle[aps]{revtex}
\draft
\input epsf
\begin{document}
\twocolumn[\hsize\textwidth\columnwidth\hsize\csname@twocolumnfalse%
\endcsname
\title{Quantum Force in Superconductor}
\author{A. \ V. \ Nikulov}

\address{Institute of Microelectronics Technology and High Purity Materials, Russian Academy of Sciences, 142432 Chernogolovka, Moscow District, RUSSIA}

\maketitle
\begin{abstract}
{Transitions between states with continuous (called as classical state) and
discrete (called as quantum state) spectrum of permitted momentum values is
considered. The persistent current can exist along the ring circumference
in the quantum state in contrast to the classical state. Therefore the
average momentum can changes at the considered transitions. In order to
describe the reiterated switching into and out the quantum state an
additional term is introduced in the classical Boltzmann transport
equation.  The force inducing the momentum change at the appearance of the
persistent current is called as quantum force. It is shown that dc
potential difference is induced on ring segments by the reiterated
switching if the dissipation force is not homogeneous along the ring
circumference. The closing of the superconducting state in the ring is
considered as real example of the transition from classical to quantum
state.}
\end{abstract}
]
\narrowtext

Superconductivity is a macroscopic quantum
phenomenon. This means that some macroscopic effects observed in
superconductors can not be described by classical mechanics. First of such
effects was discovered in 1933 year by W.Meissner and R.Ochsenfeld
\cite{Meissner}. The Meissner effect is caused by the quantization of the generalized
momentum $p = mv + (q/c)A$ of a quantum particles along a closed path $l$
\cite{London50}. Because $\int_{l}dl p = n2\pi \hbar$ the value $\int_{l}dl
A  + (mc/q)\int_{l}dl v = \Phi + (mc/q)\int_{l}dl v_{s}$ called by F.London
\cite{London50} as fluxoid is quantized: $\Phi + (mc/q)\int_{l}dl v =
\Phi_{0}n$ \cite{tink75}. $A$ is the vector potential; $\Phi = \int_{l} dl
A$ is the magnetic flux contained within the closed path $l$; $\Phi_{0} =
2\pi \hbar c/q$ is the flux quantum. The charge of the superconducting pair
$q = 2e$. Therefore $\Phi_{0} = \pi \hbar c/e$ for superconductors
\cite{Onsager}.

The magnetic flux is expelled from the interior of a superconductor because
the superconducting pairs are condensed bosons with the same $n$ value.
When the wave function of the superconducting pairs does not have any
singularity inside $l$, it can be tightened to point
without crossing of a singularity. Therefore the $n$ value and consequently
the fluxoid value should be equal zero $\Phi + (mc/e)\int_{l}dl v_{s} =
\Phi_{0}n = 0$, i.e. in the interior of the superconductor,
where $v_{s} = 0$, the magnetic flux $\Phi$ should be equal zero.

When the $l$ is a closed path along the ring circumference the $n$ can be
any integer number. When the ring wall is wide, $w \gg \lambda$, the flux
quantization $\Phi = \Phi_{0}n$ \cite{Byers} takes place because $v_{s} =
0$ in the interior of the superconductor. Here $\lambda$ is the penetration
depth of magnetic field. The quantization of the magnetic flux was observed
first in \cite{Deaver}. In a superconducting ring (or tube) with narrow
wall the velocity of superconducting pairs is quantized $$\int_{l}dl v_{s}
= \frac{\hbar}{Rm}(n - \frac{\Phi}{\Phi_{0}}) \eqno{(1)}$$ This
quantization was manifested first by the Little-Parks experiment
\cite{little}. The limit case of the ring with narrow wall ($w \ll R,
\lambda$) is considered in the present paper.

The nonzero average velocity in the thermodynamic equilibrium state means
the existence of the persistent current $j_{p.c.}$ because $j = qn_{q}<v>$
\cite{LLqm}. Here $n_{q}$ is the density of particles with charge $q$.
There is a principle difference between classical and quantum mechanics.
The persistent current is equal

$$j_{p.c.} = q\sum_{p}vf_{0} = \frac{q}{m}\sum_{p}(p - \frac{q}{c}A) f_{0}
\eqno{(2)}$$ The distribution function of the equilibrium state, $f_{0}$,
is even function of the velocity because it depends only on the relation
$E_{p}/k_{B}T$ and the kinetic energy $E_{p}$ is proportional to $v^{2}$ in
a consequence of the space symmetry: $E_{p} = mv^{2}/2 = (p -
(q/c)A)^{2}/2m$.

When the spectrum of permitted $p$ values is continuous the summation in
(2) can be replaced by the integration and $p = mv + (q/c)A$ can be
replaced by $p = mv$. $\int_{-\infty }^{\infty } dp \ pf_{0} = 0$ because
$f_{0}$ is even function of $p = mv$. Consequently, the persistent current
can exist only in states with discrete $p$ spectrum. According to the
classical mechanics all $p$ values are permitted. The discrete spectrum is
characteristic feature of the quantum mechanics. Therefore the state with
continuous $p$ spectrum will be called in this paper as classical state and
the one with discrete spectrum as quantum state. The summation in (2) can
be replaced by the integration when the energy difference between adjacent
permitted states $E_{n+1} - E_{n}$ is small in comparison with $k_{B}T$ therefore
the $p$ spectrum may be considered as continuous at $(E_{n+1} - E_{n})/k_{B}T \ll
1$.

The persistent current $j_{p.c.} = (q/m)\sum_{p} (p - q\Phi/cl)f_{0} =
(c\Phi_{0}/\lambda ^{2}l)\chi $ can exist in a ring when the $p$ spectrum
along the ring circumference $l = 2\pi R$ is discrete. Here $\lambda =
(mc^{2}/n_{ef}q^{2})^{1/2}$ is an analog of the London penetration depth;
$n_{ef}$ is an effective density; $\chi  = \chi (\Phi/\Phi_{0})$ is a
periodic function of the magnetic flux $\Phi$ contained within the ring.
$-05 < \chi < 0.5$, $\chi = 0$ at $\Phi/\Phi_{0} = n$ and $\Phi/\Phi_{0} =
n + 0.5$. In the case of a homogeneous superconducting ring $n_{ef} =
n_{s}$ is the density of supercondicting pairs and $\lambda $ is the London
penetration depth.

The persistent current can be not equal zero for all statistics:
Bose-Einstein, Fermi-Dirac and classical. But in normal metal it can
observed only in a perfect mesoscopic ring with small radius at very low
temperature. Electrons have discrete spectrum of momentum along the ring
circumference $p_{n} = (\hbar /R)n$ when it's mean free path $l_{f.p.} \gg
l$. In the opposite limit, $l_{f.p.} \ll l$, electron may be considered as
classical particles having at the same time a coordinate and a momentum
\cite{MetTheor}. The momentum uncertainty in this case $\Delta p >
\hbar/\Delta x > \hbar/l_{f.p.}$ exceeds the momentum difference between
adjacent permitted states $p_{n+1} - p_{n} = (\hbar /R)$, i.e. the spectrum
is continuous. Therefore the persistent current can exist only at $l_{f.p.}
\gg l$ and at very low temperature because for one electron $E_{n+1} -
E_{n} \simeq \hbar^{2}/R^{2}2m \simeq k_{B}T$ at $T = 1 \ K$ and $R = 0.6 \
\mu m$.

In a superconducting ring $l_{f.p.} = \infty $ and $E_{n+1} - E_{n} \simeq
sln_{s}\hbar^{2}/R^{2}2m \gg \hbar^{2}/R^{2}2m$ because superconducting
pairs are condensed bosons. $sln_{s}$ is the number of superconducting
pairs in the ring; $s$ is the area of ring wall section. Therefore superconductivity is a macroscopic quantum
phenomenon. The persistent current can exist in superconductor with big radius. In a homogeneous superconducting ring
$j_{p.c.} = 2e(\hbar/Rm)n_{s}\chi $, where $\chi \simeq n - \Phi/\Phi_{0}$
at $sln_{s}\hbar^{2}/R^{2}2m \gg k_{B}T$. Here the $n$ is conformed with
minimum permitted velocity, i.e. $-0.5 < n - \Phi/\Phi_{0} < 0.5$
\cite{tink75}.

The possibility of the persistent current was pointed out first for the
Bose-Einstein condensation \cite{Blatt}. Such current is observed both in
superconductors and superfluids (see \cite{Tilley}). The rotation of the
reservoir with superfluid takes the place of the magnetic flux $\Phi$ at $q
= 0$ \cite{Tilley}. The persistent current of electrons in normal metal
mesoscopic systems was predicted first by Kulik \cite{Kulik} and later
rediscovered by Buttiker et al. \cite{Buttiker}. This predictions was
confirmed by experimental result \cite{IBM1991}. Now this phenomenon is in
progress to study in different aspects \cite{Zipper}.

The division of classical and quantum states used in this paper is
relative. For example, quantum effects, such as the interference effect
\cite{Sharvin}, are observed at short elastic mean free path and long
inelastic mean free path. Although the $p$ spectrum is continuous at short
elastic mean free path. The persistent current is reduced with the increase
of both inelastic and elastic scattering \cite{Landauer}.

Thus, according to our modern knowledge the equilibrium states of a
mesoscopic ring may be distinguished by such macroscopic parameter as the
persistent current: $\sum_{p} qvf_{cl} = 0$ in the classical state with
continuous $p$ spectrum and $\sum_{p} qvf_{qu} = j_{p.c.}$ in the quantum
state with discrete $p$ spectrum. Here $f_{cl}$ is the distribution
function of the equilibrium classical state and $f_{qu}$ is the
distribution function of the equilibrium quantum state. The object of the
present paper is the consideration of reiterated transitions between
$f_{cl}$ and $f_{qu}$.

The relaxation of the persistent current after the transition to the
classical state (from $f_{qu}$ to $f_{cl}$) can be described by the
Boltzmann transport equation \cite{MetTheor} $$\frac{df}{dt} =
\frac{\partial f}{\partial t} + v \frac{\partial f}{\partial l} + qE
\frac{\partial f}{\partial p} = -\frac{f_{1}}{\tau} \eqno{(3)} $$ $\tau $
is the mean time between collisions; $f_{1} = f - f_{0}$ is the deviation
of the distribution function from the equilibrium state ($f_{0} = f_{cl}$
in this case). One-dimensional case is considered: $f$ changes only along
the ring circumference.

In order to describe this process we do not must exceed the limits of the
classical mechanics. The disappearance of a current in a consequence of the
dissipation is usual process of the classical mechanics. Some assumptions
must be made in the derivation of the transport equation (3) \cite{Kohn57}.
First of all it is the "random phase" assumption. It is assumed these
assumptions are valid in the classical state.

The appearance of the persistent current contradicts to the classical
mechanics. But the transport equation can be used for the phenomenological
description of the transition from $f_{cl}$ to $f_{qu}$ if a new term
$\aleph $ is added to it:

$$\frac{df}{dt} = \frac{\partial f}{\partial t} + v \frac{\partial
f}{\partial l} + qE \frac{\partial f}{\partial p} = \sum_{t_{q}} \aleph (t -
t_{q}) - \frac{f_{1}} {\tau} \eqno{(4)} $$ $\aleph (t - t_{q})$ is not
equal zero only during a time interval from $t = t_{q}$ to $t = t_{q} +
\Delta t_{q}$ when the transition from the classical to quantum equilibrium
state takes place. According to (4) $\int_{\Delta t_{q}} dt \aleph =
\int_{\Delta t_{q}} dt df/dt + \int_{\Delta t_{q}} dt f_{1}/\tau = f_{qu} -
f_{cl} + \int_{\Delta t_{q}} dt f_{1}/\tau $.

The generalized momentum along the ring circumference changes on $\Delta P
= (m/q)j_{p.c.}[1 + (L/l)(s/\lambda^{2})]$ at the transition between the
classical and quantum states. According to the classical mechanics any
momentum change is caused by a force. The balance on the average forces

$$\frac{\partial P}{\partial t} - F_{p} - F_{e} = \sum_{t_{q}} F_{q}(t - t_{q})
- F_{dis} \eqno{(5)} $$ is obtained by multiplication of the transport
equation (4) by the momentum and summing up the $p$ states. $P =
\sum_{p}pf$; $F_{p} = - \partial (\sum_{p}pvf)/\partial l = - \partial
(n_{q}<pv>)/\partial l$ is the force of the pressure; $F_{e} = - qE\sum_{p}
p \partial f/\partial p = qEn_{q}$ is the force of the electric field;
$F_{dis} = \sum_{p} p f_{1}/\tau $ is the dissipation force. $F_{q} =
\sum_{p} p \aleph$. $\int_{\Delta t_{q}} dt F_{q} = \int_{\Delta t_{q}} dt
\sum_{p} p \aleph = \sum_{p} p f_{qu} - \sum_{p} p f_{cl} + \int_{\Delta
t_{q}} dt \sum_{p} p f_{1}/\tau = (m/q)j_{p.c.}[1 + (L/l)(s/\lambda^{2})] +
\int_{\Delta t_{q}} dt F_{dis}$. The $F_{q}$ describes the momentum change
at the transition to the quantum state. Therefore it is natural to call it
as quantum force.

After the full cycle: $f_{cl} \rightarrow f_{qu} \rightarrow f_{cl}$, the
momentum $P$ remains invariable, $\int_{t_{cyc}}dt\partial P/\partial t =
0$. Consequently, according to (5) $\int_{t_{cyc}}dt (F_{p} + F_{e} + F_{q}
- F_{dis}) = 0$. $\int_{l}dl F_{p} = - \int_{l}dl \partial
(n_{q}<pv>)/\partial l = 0$; $\int_{t_{cyc}}dt \int_{l}dl F_{e} = qn_{q}
\int_{t_{cyc}}dt\int_{l}dl E = -(qn_{q}/c) \int_{t_{cyc}} dt d\Phi/dt =
(qn_{q}/c) (\Phi - \Phi) = 0$. Consequently $\int_{t_{cyc}}dt \int_{l}dl
F_{q} = \int_{t_{cyc}}dt \int_{l}dl F_{dis}$.

Thus, the quantum force is opposed to the dissipation force. There is
important difference between the quantum and dissipation forces in a
consequence of the uncertainty in space. The quantum force can not be
localized in time and in space in a consequence of the uncertainty
relations: $\Delta E \Delta t > \hbar$ and $\Delta p \Delta l > \hbar$. The
uncertainty in time is not important here because the dissipation force is
the average value of the forces acting at chaotic collisions. But the
dissipation force can be localized in space. Whereas the quantum force can
not be localized along the ring circumference in principle because a
particle should be spread all over ring in the quantum state: $ \Delta l >
\hbar/\Delta p \gg \hbar/(p_{n+1} - p_{n}) = R$.

The dissipation $\int_{\Delta t_{q}} dt F_{dis}$ during the transition from
$f_{cl}$ to $f_{qu}$ is weak when the time $\Delta t_{q}$ is long. In this
case (which is considered in the present paper) the quantum force
$\int_{\Delta t_{q}} dt F_{q} \simeq (m/q)j_{p.c.}[1 +
(L/l)(s/\lambda^{2})]$ is constant along the ring circumference because
$I_{p.c.} = sj_{p.c.}$ should be constant in the equilibrium state (it is
proposed $s$ is constant). Therefore, $\int_{t_{cyc}}dt
F_{q} - \int_{t_{cyc}}dt F_{dis} \neq 0$ and consequently $\int_{t_{cyc}}dt
F_{p} + \int_{t_{cyc}}dt F_{e} \neq 0$ in an inhomogeneous ring, in which
$F_{dis}$ is not constant along the ring circumference.

Both $F_{p}$ and $F_{e}$ are induced by deviation of the electron density
$\Delta n_{q}$ from the equilibrium value ($\Delta n_{q} \ll n_{q}$). In
the order of value $F_{p} \approx - <pv>\Delta n_{q}/\Delta l$ and $F_{e}
\approx q^{2} n_{q} \Delta n_{q}\Delta l = q^{2}/n_{q}^{-1/3} (\Delta
l/n_{q}^{-1/3})^{2} \Delta n_{q}/\Delta l$. The characteristic length
$\Delta l$ of $\Delta n_{q}$ change is much longer than the distance
between electrons: $\Delta l \gg n_{q}^{-1/3}$. In any metal $<pv> \approx
q^{2}/n_{q}^{-1/3}$ \cite{MetTheor}. Consequently, $F_{p} \ll F_{e}$ and
$\int_{t_{cyc}}dt F_{e} \simeq  \int_{t_{cyc}}dt F_{dis} - \int_{t_{cyc}}dt
F_{q}$

This means that the reiterated switching into and out the quantum state can induced the voltage
in segments of inhomogeneous ring the average value of which by a long time $T_{long}$, $<E> \simeq (<F_{dis}> - <F_{q}>)/qn_{q} $, is not equal zero. The dc potential difference on a segment $l_{a}$ is equal $V_{a} = l_{a}<E> \simeq - l_{a}<F_{q}>/qn_{q}$ when $<F_{dis}> =
0$ in this segment. $<F_{q}> = (\int_{T_{long}}dt F_{q})/T_{long} =
(\sum_{t_{q}} (m/q)j_{p.c.}(q)[1 + (L/l)(s/\lambda^{2})])/T_{long} =
(m/q)<j_{p.c.}>[1 + (L/l)(s/\lambda^{2})]f$. Consequently, $$V_{a} =  -
\frac{l_{a}}{l}\frac{<n_{ef}>}{n_{q}}\frac{\Phi_{0}f}{c}(1 +
\frac{L}{l}\frac{s}{\lambda^{2}}) \chi \eqno{(6)}$$ $<j_{p.c.}> = \sum_{t_{q}}
j_{p.c.}/N_{l}$ is the average persistent current in the quantum states; $f
= N_{l}/T_{long}$ is the average frequency of the switching; $N_{l}$ is the
number of the switching during the time $T_{long}$. The potential
difference on other segment $l_{b}$, $V_{b} = - V_{a}$ because $\int_{l} dl
<E> = 0$, ($l_{a} + l_{b} =l$).

The closing of the superconducting state in the ring is the real example of
the transition from classical to quantum state. The superconducting current
$j_{s} = 2en_{s}v_{s}$ should be constant along the ring circumference in
the equilibrium state. Therefore the kinetic energy of superconducting pair
in the ring is equal $E_{p} = s\int_{l}dl n_{s}2mv_{s}^{2}/2 =
(sj_{s}/4e)\int_{l}dl 2mv_{s} = (sj_{s}2\pi \hbar/4e)(n - \Phi/\Phi_{0})$
(here the velocity quantization (1) is taken into account). According to
(1) $j_{s}  = (e\hbar/Rm<n_{s}^{-1}>)(n - \Phi/\Phi_{0})$, where
$<n_{s}^{-1}> = l^{-1}\int_{l}dl n_{s}^{-1}$. Consequently, $E_{p} =
(s2\pi \hbar^{2}/4Rm<n_{s}^{-1}>)(n - \Phi/\Phi_{0})^{2}$ and the energy of
the magnetic flux induced by the superconducting current $E_{L} =
LI_{s}^{2}/2c^{2} =
(Ls^{2}e^{2}\hbar^{2}/2c^{2}R^{2}m^{2}<n_{s}^{-1}>^{2})(n -
\Phi/\Phi_{0})^{2} = (Ls/l\lambda _{0}^{2})n'_{s}E_{p}$. Here
$\lambda _{0} = (c^{2}2m/4e^{2}n_{s}(0))^{1/2}$ is the London penetration
depth at $T = 0$; $n'_{s} = (n_{s}(0)<n_{s}^{-1}>)^{-1}$; $n_{s}(0)$ is the density of superconducting pairs at $T
= 0$. 

In a strongly inhomogeneous ring in which the density $n_{s}$ in a segment
$l_{b}$ is much lower that in other segment $l_{a}$, $<n_{s}^{-1}> \simeq
l^{-1}\int_{l_{b}}dl n_{s}^{-1} \approx l_{b}/ln_{sb}$. There can be taken
into account the Josephson current if $n_{sb}$ is considered as an
effective density. The energy difference between adjacent permitted states
$E_{n+1} - E_{n} \simeq s2\pi \hbar^{2}/4Rm<n_{s}^{-1}> \approx s\pi^{2}
\hbar^{2}n_{sb}/l_{b}m$ is determined by lowest density $n_{sb}$. Therefore
the $p$ spectrum may be considered as continuous when the superconducting
state in the ring is not closed, i.e. when $n_{sb} \approx 0$ and
consequently $(E_{n+1} - E_{n})/k_{B}T \ll 1$. The "random phase"
assumption is valid in this case. At large $n_{sb}$ value $(E_{n+1} -
E_{n})/k_{B}T \gg 1$. Thus, the persistent current appears at the closing
of the superconducting state and disappears at the break of phase coherence
along the ring circumference.

The quantum force accelerates the superconducting pair in the $l_{a}$
segment from zero to $v_{s} =  (2\pi \hbar
/m)(n_{sb}/(l_{a}n_{sb}+l_{b}n_{sa}))(n - \Phi/\Phi_{0}) $ against the
force of the electric field $F_{e} = - n_{sa}(2e/c)LdI/dt$. Here $n_{sa}$
and $n_{sb}$ are average density of superconducting pairs in the segment
$l_{a}$ and $l_{b}$. The electric force acts both on superconducting pairs
and normal electrons. Whereas the quantum force acts only on
superconducting pairs and the dissipation force acts only on normal
electrons. Therefore the balance on the average forces $<F_{e}> =
2en_{sa}<E> + en_{e}<E> \simeq <F_{dis}> - <F_{q}> $ falls apart into the
balance of forces acting on superconducting pairs $2en_{sa}<E> \simeq -
<F_{q}> $ and on normal electrons $en_{e}<E> \simeq <F_{dis}> $. It is
assumed that the densities of superconducting pairs $n_{sa}$ and normal
electrons $n_{e}$ in the $l_{a}$ segment are not change.

At $(E_{n+1} - E_{n})/k_{B}T \gg 1$, $j_{p.c.} = j_{s} =  (4e\pi \hbar
/m)(n_{sa}n_{sb}/(l_{a}n_{sb}+l_{b}n_{sa}))(n - \Phi/\Phi_{0}) $ with the
$n$ value conformed with minimum permitted velocity, i.e. $-0.5 < n -
\Phi/\Phi_{0} < 0.5$. At $(E_{n+1} - E_{n})/k_{B}T \ll 1$, $j_{p.c.} = 0$.
Consequently, the dc potential difference $V_{a} =
(l_{a}n_{sbm}/(l_{a}n_{sbm}+l_{b}n_{sa})) [1 + (L/l)(s/\lambda^{2})] \chi
\Phi_{0}f/c$ is induced on the superconducting segment $l_{a}$ by the
reiterated switching of the $n_{sb}$ from a large value $n_{sbm}$ to
$n_{sb} \approx 0$ with the average frequency $f$.

Thus, the inhomogeneous superconducting ring
can be considered as dc generator. When the $n_{sb}$ change is induced by
temperature change then the heat energy is transformed in electric energy.
The superconducting state can be also interrupted and closed by mechanical
interrupting and closing of the ring. In this case the mechanical energy
can be transformed to the electric energy. The superconducting ring can not
substitute traditional generators because the voltage induced in it is very
weak. Because $l_{a}n_{sbm}/(l_{a}n_{sbm}+l_{b}n_{sa}) < 1$, $1 +
(L/l)(s/\lambda^{2}) \approx 1$ and $|\chi | < 0.5$ the maximum voltage
$V_{a} \simeq  \Phi_{0}f/2c \simeq 10^{-9}f \ \mu V$. But the consideration
of this energy transformation gives important physical results.

According to the classical mechanics, in order to change energy value a
work should be done. The work done by the quantum force produces the
kinetic $E_{p}$ and magnetic flux $E_{L}$ energy: $$E_{p} + E_{L} =
s\frac{\pi \hbar^{2}}{2Rm}\frac{1}{<n_{s}^{-1}>}(1 +
\frac{L}{l}\frac{s}{\lambda_{0}^{2}}n'_{s}) (n -
\frac{\Phi}{\Phi_{0}})^{2} \eqno{(7)}$$. Both these energies increase the
energy of the superconducting state at $\Phi/\Phi_{0} \neq n$. The
Little-Parks experiment \cite{little} is explained \cite{tinkham} by the
periodical dependence $E_{p}$ on $\Phi$. The energy of magnetic flux does
not influence on the value of the critical temperature because the $E_{L}$
is proportional to $n_{s}^{2}$ ($<n_{s}^{-1}> = 1/n_{s}$ in a homogeneous
ring). The Little-Parks effect in an inhomogeneous superconducting ring is
considered in \cite{jltp98}.

We may say that the critical temperature is reduced at $\Phi /\Phi_{0} \neq
n$ because the quantum force should be overcome at the
transition to the superconducting state. This explanation does not differ
from the Tinkham's explanation \cite{tinkham} and does not give any new knowledge because the critical temperature is scalar. The introducing of the quantum force is useful at the consideration of vector quantities, such as the voltage.

The work $W$ is the product of a force $F$ and a path $dx$: $W = \int dx
F$. It is impossible to point out a path when the closing of the
superconducting state is induced by temperature change. But the real path
exists when the mechanical closing of the superconducting ring interrupted
by Josephson junction takes place. The Josephson current decreases
exponentially with increasing of break width $b$ and has ignorable value
when $b$ exceeds some nanometers \cite{Barone}. Therefore the energy
$E_{p} + E_{L}$ increases from zero to the value determined by the relation
(7), $E_{p} + E_{L} \approx E_{p} \approx (s/\lambda
^{2})(\Phi_{0}^{2}/4\pi R) (n - \Phi /\Phi_{0})^{2}$, when the $b$ value is
changed from some nanometers to zero. This means: in order to close the
ring the work $W \simeq E_{p} + E_{L}$ should be expended, i.e. the quantum
force, the average value of which is equal $W/\Delta b$ should be overcome.
At $s/\lambda ^{2} \simeq 1$, $n - \Phi /\Phi_{0} = 1/2$ and $R = 1 \mu m$,
$E_{p} \approx 3 \ 10^{-20} \ J$. Consequently, at $\Delta b = 10 \ nm$,
$W/\Delta b \approx 3 \ 10^{-20} \ N$. It is very weak force. But there is
important to show a connection of the quantum force with a real classical
force. This consideration shows that the wave function can have an
elasticity.

\end{document}